%% file: Template.tex
\documentclass{article}
\usepackage{lmodern}
\usepackage{microtype}
\usepackage{spconf,graphicx,hyperref}
\usepackage{amsmath,amssymb,amsfonts}
\usepackage{bm}

\usepackage{booktabs}
\usepackage{multirow}
\usepackage{makecell}
\usepackage{graphicx}
\usepackage{float}
\usepackage{subcaption}
\usepackage{array}
\usepackage{colortbl}
\usepackage{tabularx}
\usepackage{wrapfig}

\title{AURA: Unified Multimodal Framework for Conversational Music Editing}
\name{Quoc-Huy Trinh$^{1,4}$ \qquad Minh-Van Nguyen$^{2,4}$ \qquad Debesh Jha$^{3}$}
\address{$^{1}$Aalto University \quad $^{2}$Technical University of Denmark \quad$^{3}$University of South Dakota \quad $^{4}$OpenRB Lab}
\begin{document}
%
\maketitle
\begin{abstract}
Instruction-guided music editors typically process each request independently, limiting their ability to support workflows in which users progressively refine a track. We introduce AURA, a unified multimodal framework for conversational music editing. AURA uses a multimodal large language model to interpret the complete dialogue history, an optional image, and reference audio, distilling the editing intent into compact concept tokens. A concept-to-audio module injects these tokens and frame-aligned reference features into a frozen MusicGen backbone, enabling precise edits while preserving unaffected content. AURA optimizes only 91M parameters while retaining 1.9B frozen backbone parameters. Experiments on Slakh2100 and MoisesDB demonstrate substantial improvements in edit correctness and content preservation over existing instruction-guided methods, including a \(4\)--\(5\times\) reduction in FAD for out-of-domain addition and removal. \url{https://openrb-lab.github.io/AURA-demo/}

\end{abstract}
\input{sections/1introduction}

\input{sections/2method}

\input{sections/3experimental}
\input{sections/4conclusion}

\small
\bibliographystyle{IEEEbib}
\bibliography{strings,refs}

\end{document}

%% file: sections/1introduction.tex
\section{Introduction}

Recent advances in text-to-music generation have substantially expanded
the possibilities for AI-assisted composition and production. Recent works~\cite{agostinelli2023musiclm,melechovsky2024mustango,stableaudio,musicgen,evans2025stable, evans2026stable,ning2025diffrhythm,jiang2025diffrhythm,liu2025songgen} generate realistic
music from textual descriptions using autoregressive or diffusion-based
models. This progress has motivated a shift from generation toward
controllable music editing, where users modify an existing track through
natural-language or visual instructions while preserving its unaffected
content.

Recent methods, including AUDIT~\cite{wang2023audit},
InstructME~\cite{han2023instructme},
M$^2$UGen~\cite{liu2023m},
Instruct-MusicGen~\cite{zhang2024instruct}, and
LeVo 2~\cite{lei2026levo}, support instruction-guided editing and
stem-level operations such as adding, removing, or extracting musical
sources. However, these systems generally process each request
independently and therefore cannot resolve editing instructions that
depend on earlier dialogue turns. This limitation is particularly
restrictive in realistic production workflows, where users progressively
refine a track and expect each edit to remain consistent with the evolving
conversation.

To address this limitation, we introduce AURA, a unified multimodal
framework for conversational music editing. AURA uses a multimodal large
language model to jointly interpret the complete dialogue history, an
optional image, and reference audio. It distills the editing
intent into concept tokens, which are projected into the
conditioning space of a frozen MusicGen backbone. A concept-to-audio
(C2A) module is proposed to combine these semantic representations with
reference-music features, enabling the decoder to apply the
requested edit while preserving musicality prior. Only the LoRA
adapters~\cite{hu2021lora}, projectors, and C2A modules are trainable, retaining the
pretrained generative prior of the backbone.

In summary, our main contributions are:
\begin{itemize}
    \item We propose concept-guided music decoding, which represents
    multimodal editing intent using compact concept tokens and injects
    them into a pretrained music decoder through a C2A module. The module
    combines semantic edit control with frame-aligned reference
    conditioning to preserve unedited content.

    \item We introduce AURA, a unified multimodal framework that jointly
    reasons over dialogue history, images, and reference audio for
    coherent multi-turn music editing. AURA substantially improves edit
    correctness and content preservation across in-domain and
    out-of-domain.
\end{itemize}

We train AURA on agent-generated editing conversations constructed from
the Slakh2100 training set~\cite{manilow2019cutting} and evaluate it on
Slakh2100-test as an in-domain benchmark and
MoisesDB~\cite{pereira2023moisesdb} as an out-of-domain benchmark.
Experiments on single- and multi-turn editing demonstrate strong
generalization and substantially better content preservation than
existing instruction-guided approaches.

%% file: sections/2method.tex
\vspace{-2mm}
\section{Method}
\begin{figure*}[!ht]
    \centering
    \includegraphics[width=0.8\linewidth]{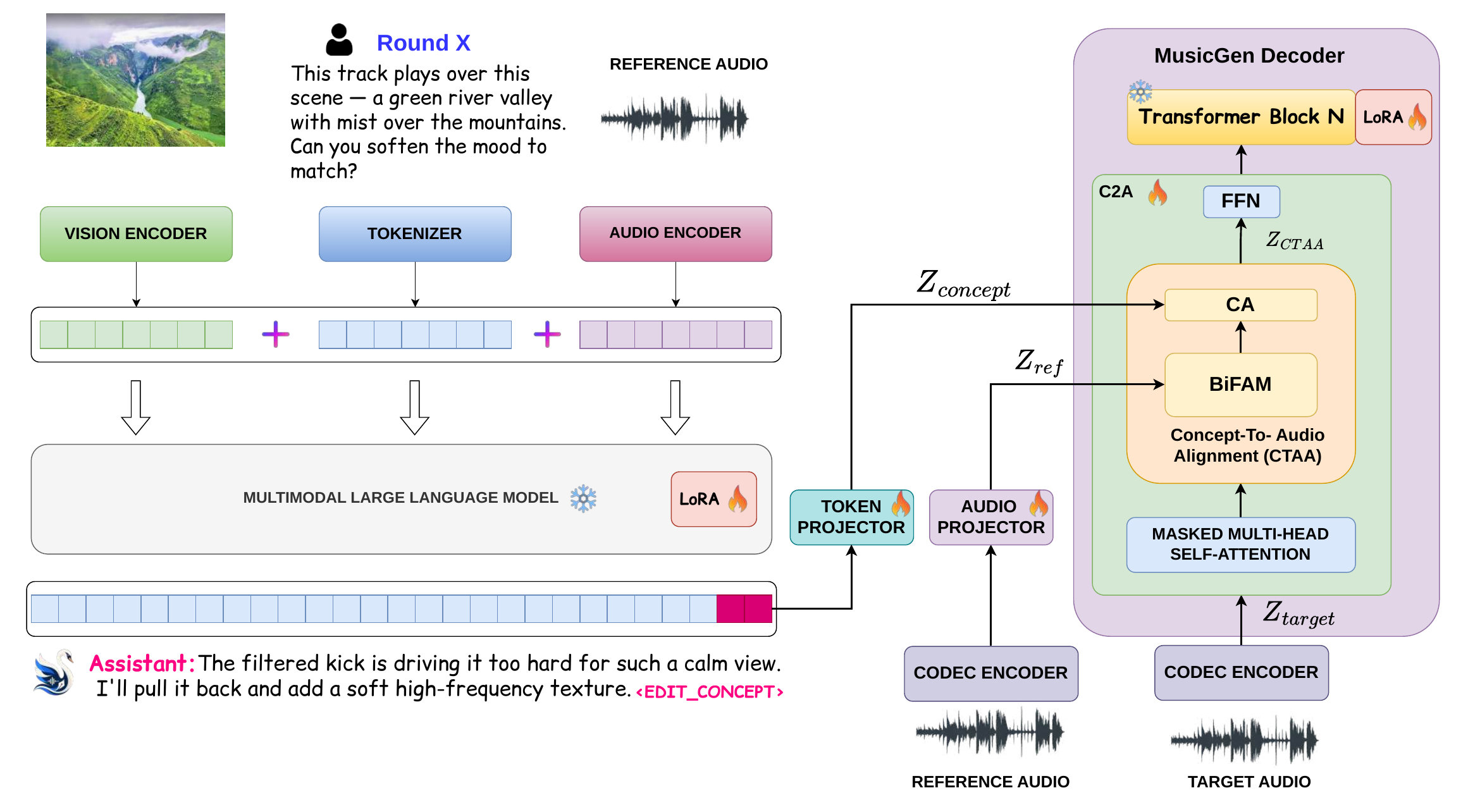}
    \caption{\textbf{Overall architecture of AURA.} Given an image $I$, a
    multi-turn conversation $T$, and a reference audio $X_{\mathrm{ref}}$,
    the multimodal LLM $g_{\mathrm{mllm}}$ produces a natural-language
    response together with $\langle\mathrm{EDIT\_CONCEPT}\rangle$ tokens
    that summarize the editing intent inferred from all modalities and the
    dialogue history. The projector $g_{\mathrm{proj}}$ maps the hidden
    states of these tokens into the music decoder's conditioning space.
    The concept-to-audio module $g_{\mathrm{c2a}}$ fuses this concept
    embedding, together with the reference audio stream, into the hidden
    states of the frozen music decoder, which then generates the edited
    audio $X_{\mathrm{target}}$.}
    \label{fig:aural_overall}
\end{figure*}

\subsection{Overall framework}

Given an optional image $I$, a multi-turn conversation history
$\mathcal{T}$, and a reference audio signal $X_{\mathrm{ref}}$, AURA
generates an edited audio signal $\hat{X}_{\mathrm{tgt}}$:
\begin{equation}
    \hat{X}_{\mathrm{tgt}}
    =
    g\!\left(I,\mathcal{T},X_{\mathrm{ref}}\right).
    \label{eq:overall}
\end{equation}
AURA separates multimodal intent understanding from audio generation.
A multimodal large language model (MLLM) infers the desired edit from
the complete conversation and input modalities, while a MusicGen-based
decoder applies the edit and preserves the unaffected content of the
reference audio. As illustrated in Fig.~\ref{fig:aural_overall}, AURA comprises an MLLM, concept and audio projectors, a concept-to-audio (C2A) module plugged in the
MusicGen decoder. The pretrained MLLM and MusicGen backbone parameters
remain frozen, while their LoRA adapters, the projectors, and the C2A
module are optimized during training.

\subsection{Multimodal large language model}

Prior music-editing models~\cite{liu2023m,zhang2024instruct,han2023instructme, lei2025levo}
only use the current instruction and reference audio but
do not allow dialogue input, limiting them to single-turn interaction.
To support conversational editing, we employ the Thinker component of
Qwen2.5-Omni~\cite{xu2025qwen25omnitechnicalreport} and introduce
$n_c$ instances of a special
$\langle\mathrm{EDIT\_CONCEPT}\rangle$ token. The MLLM jointly processes
the conversation history $\mathcal{T}$, image $I$, and reference audio
$X_{\mathrm{ref}}$:
\begin{equation}
    Y
    =
    g_{\mathrm{mllm}}
    \!\left(I,\mathcal{T},X_{\mathrm{ref}}\right),
    \label{eq:mllm}
\end{equation}
where $Y$ is a natural-language response optionally followed by the
concept-token block. Because these tokens attend to all preceding
modalities and dialogue turns, their final hidden states compactly encode
the editing operation, target instrument, desired attributes, and
conversational constraints. We denote these states by
\begin{equation}
    \mathbf{H}_{\mathrm{concept}}
    \in
    \mathbb{R}^{n_c\times d_{\mathrm{llm}}},
    \label{eq:concept_hidden}
\end{equation}
where $d_{\mathrm{llm}}=3584$. The concept states provide a compact
semantic interface between multimodal understanding and audio generation.
When no audio edit is required, the MLLM produces only a textual response
and emits no concept tokens.

\subsection{Concept and audio projectors}

The concept and reference-audio representations originate from different
feature spaces and must therefore be aligned with the MusicGen hidden
space. For the concept tokens, we apply a LayerNorm followed by a
two-layer MLP:
\begin{equation}
    \mathbf{Z}_{\mathrm{concept}}
    =
    \mathbf{W}_2
    \,\mathrm{GELU}\!\left(
        \mathbf{W}_1
        \,\mathrm{LN}\!\left(
            \mathbf{H}_{\mathrm{concept}}
        \right)
    \right),
    \label{eq:concept_projector}
\end{equation}
where
$\mathbf{Z}_{\mathrm{concept}}
\in\mathbb{R}^{n_c\times d_{\mathrm{m}}}$
and $d_{\mathrm{m}}$ is the MusicGen hidden dimension. Layer normalization
reduces sensitivity to changes in the scale of the MLLM hidden states
during post-training. The resulting $n_c$ vectors serve as the keys and
values of the concept cross-attention in each decoder layer.

In parallel, the frozen EnCodec encoder extracts a frame-level
representation of the reference audio. The audio projector maps this
representation into the same hidden space:
\begin{equation}
    \mathbf{Z}_{\mathrm{ref}}
    =
    g_{\mathrm{aud}}\!\left(
        E_{\mathrm{codec}}(X_{\mathrm{ref}})
    \right)
    \in
    \mathbb{R}^{L\times d_{\mathrm{m}}},
    \label{eq:audio_projector}
\end{equation}
where $L$ denotes the number of audio frames. The two projected
representations have primary roles:
$\mathbf{Z}_{\mathrm{concept}}$ specifies what should be changed, whereas
$\mathbf{Z}_{\mathrm{ref}}$ provides frame-aligned information about the
content that should be preserved.

\subsection{Music decoder}

The music decoder is based on MusicGen-medium~\cite{musicgen}, which
autoregressively predicts four streams of EnCodec~\cite{defossez2022high} codes at 50\,Hz. Its
pretrained backbone remains frozen to preserve the learned musical prior.
Editing is introduced through trainable LoRA adapters and the proposed
C2A module, which integrates the reference-audio representation
$\mathbf{Z}_{\mathrm{ref}}$ with the edit representation
$\mathbf{Z}_{\mathrm{concept}}$.

\noindent\textbf{Frame-aligned reference fusion.}
The reference audio forms a generation-independent stream that is aligned
with the autoregressive music stream. Let
$\mathbf{H}_{r}^{(\ell)}$ and $\mathbf{H}_{m}^{(\ell)}$ denote the
reference and music states, respectively, at decoder layer $\ell$.
The two streams use the same frozen attention projections:
\begin{equation}
(\mathbf{Q}_{x}^{(\ell)}, \mathbf{K}_{x}^{(\ell)}, \mathbf{V}_{x}^{(\ell)})
= \mathbf{H}_{x}^{(\ell)}
(\mathbf{W}_{Q}^{(\ell)}, \mathbf{W}_{K}^{(\ell)}, \mathbf{W}_{V}^{(\ell)}),
\quad x\in\{r,m\},
\label{eq:stream_projections}
\end{equation}
Because the reference stream does not attend to the generated sequence,
its representations can be computed once and cached throughout
autoregressive decoding. Our \textbf{bidirectional feature-alignment module
(BiFAM)} uses a shared, position-aligned query to retrieve information
from both streams:
\begin{equation}
\begin{aligned}
    \mathbf{S}^{(\ell)}
    &=
    \alpha_{r}^{(\ell)}
    \,\mathrm{Attn}\!\left(
        \mathbf{Q}_{r}^{(\ell)}+\mathbf{Q}_{m}^{(\ell)},
        \mathbf{K}_{r}^{(\ell)},
        \mathbf{V}_{r}^{(\ell)};
        \mathcal{M}_{\mathrm{full}}
    \right) \\
    &\quad+
    \alpha_{m}^{(\ell)}
    \,\mathrm{Attn}\!\left(
        \mathbf{Q}_{r}^{(\ell)}+\mathbf{Q}_{m}^{(\ell)},
        \mathbf{K}_{m}^{(\ell)},
        \mathbf{V}_{m}^{(\ell)};
        \mathcal{M}_{\mathrm{causal}}
    \right),
\end{aligned}
\label{eq:bifam_attention}
\end{equation}
where $\mathcal{M}_{\mathrm{full}}$ and
$\mathcal{M}_{\mathrm{causal}}$ denote full and causal attention masks; $\alpha_{r}^{(\ell)}$, and
$\alpha_{m}^{(\ell)}$ are learned layer-wise weights.

Let $\mathbf{O}_{m}^{(\ell)}$ denote the decoder masked
self-attention output. BiFAM modulates this output through a FiLM
transformation:
\begin{equation}
\begin{aligned}
    \widetilde{\mathbf{O}}_{m}^{(\ell)}
    &=
    \mathbf{O}_{m}^{(\ell)}
    \odot
    \left[
        \mathbf{1}
        +
        \tanh\!\left(g^{(\ell)}\right)
        \boldsymbol{\gamma}^{(\ell)}
        \!\left(\mathbf{S}^{(\ell)}\right)
    \right] \\
    &\quad+
    \tanh\!\left(g^{(\ell)}\right)
    \boldsymbol{\beta}^{(\ell)}
    \!\left(\mathbf{S}^{(\ell)}\right),
\end{aligned}
\label{eq:bifam_film}
\end{equation}
where $g^{(\ell)}$ is a learned gate and
$[\boldsymbol{\gamma}^{(\ell)};\boldsymbol{\beta}^{(\ell)}]$ is produced
by a rank-256 bottleneck MLP. The MLP output and gate are initialized to
zero, making the transformation an identity mapping at initialization.
This frame-aligned modulation preserves the reference structure without
appending additional tokens to the autoregressive sequence.

\noindent\textbf{Concept cross-attention.}
After reference fusion, the projected concept tokens are introduced as a
shared semantic memory. At layer $\ell$, the decoder attends to
$\mathbf{Z}_{\mathrm{concept}}$ as
\begin{equation}
\begin{aligned}
    \mathbf{C}_{m}^{(\ell)}
    &=
    \mathrm{Attn}\!\left(
        \mathbf{Q}_{c}^{(\ell)}
            \!\left(\widetilde{\mathbf{O}}_{m}^{(\ell)}\right),
        \mathbf{K}_{c}^{(\ell)}
            \!\left(\mathbf{Z}_{\mathrm{concept}}\right),
        \mathbf{V}_{c}^{(\ell)}
            \!\left(\mathbf{Z}_{\mathrm{concept}}\right)
    \right), \\
    \widehat{\mathbf{O}}_{m}^{(\ell)}
    &=
    \widetilde{\mathbf{O}}_{m}^{(\ell)}
    +
    \mathbf{C}_{m}^{(\ell)}.
\end{aligned}
\label{eq:concept_cross_attention}
\end{equation}
The same $n_c$ concept vectors are shared across all decoder layers and
generation steps. Rank-64 LoRA adapters are applied only to the key and
value projections of the concept cross-attention. The resulting states
$\widehat{\mathbf{O}}_{m}^{(\ell)}$ are subsequently processed by the
layer's feed-forward network.

The zero-initialized BiFAM components and LoRA adapters preserve the
behavior of the pretrained MusicGen model at initialization, allowing
AURA to learn an editing residual rather than relearning audio generation.
Overall, AURA contains 91M trainable parameters, while 1.9B backbone
parameters remain frozen. At inference, classifier-free guidance performs
an unconditional pass by replacing
$\mathbf{Z}_{\mathrm{concept}}$ with zero-valued memory while retaining
$\mathbf{Z}_{\mathrm{ref}}$. The resulting guidance direction therefore
isolates the requested edit while holding the reference content fixed.
\vspace{-2mm}
\subsection{Objective function}
AURA is trained end-to-end with a joint objective combining the language-modeling loss over assistant responses and the music decoder's codebook cross-entropy:
\begin{equation}
\mathcal{L} = \mathcal{L}_{\mathrm{MLLM}} + \mathcal{L}_{\mathrm{music}}.
\label{equation:loss}
\end{equation}
The language loss supervises multi-turn reasoning and the placement of $\langle\mathrm{EDIT\_CONCEPT}\rangle$ tokens, while the music loss shapes their representations through the projector. This joint supervision ensures that the learned tokens are both linguistically meaningful and useful for audio generation.

%% file: sections/3experimental.tex
\vspace{-2mm}
\section{Experimental Details}
\subsection{Experimental setup}
\begin{table*}[!ht]
\footnotesize
\centering
\setlength{\tabcolsep}{3.5pt}
\caption{Comparison with SOTA music-editing methods, in-domain
(Slakh2100-test) and out-of-domain (MoisesDB). SI-SDR is undefined
for \textit{add}. AURA substantially improves edit correctness and content preservation across both benchmarks.}
\label{tab:baselines}
\begin{tabular}{ll|ccccccc|ccccccc}
\toprule
& & \multicolumn{7}{c|}{\textbf{Slakh2100-test (in-domain)}}
  & \multicolumn{7}{c}{\textbf{MoisesDB (out-of-domain)}} \\
\cmidrule(lr){3-9}\cmidrule(lr){10-16}
\textbf{Task} & \textbf{Model}
 & FAD$\downarrow$ & CLAP$\uparrow$ & KL$\downarrow$ & SSIM$\uparrow$ & P-Dem.$\uparrow$ & SI-SDR$\uparrow$ & SDRi$\uparrow$
 & FAD$\downarrow$ & CLAP$\uparrow$ & KL$\downarrow$ & SSIM$\uparrow$ & P-Dem.$\uparrow$ & SI-SDR$\uparrow$ & SDRi$\uparrow$ \\
\midrule
\multirow{4}{*}{Add}
 & M2UGen            & 4.25 & \textbf{0.29} & 1.19 & 0.08 & 0.38 & -- & --
                     & 4.64 & 0.28 & 1.04 & 0.08 & 0.45 & -- & -- \\
 & Instruct-MG       & 2.01 & 0.28 & 0.65 & 0.37 & 0.48 & -- & --
                     & 3.84 & 0.18 & 0.84 & 0.32 & 0.51 & -- & -- \\
 & LeVo              & 2.30 & 0.28 & 0.82 & 0.11 & 0.35 & -- & --
                     & 3.92 & 0.25 & 0.87 & 0.10 & 0.44 & -- & -- \\
 & \textbf{AURA}     & \textbf{0.52} & \textbf{0.29} & \textbf{0.23} & \textbf{0.78} & \textbf{0.63} & -- & --
                     & \textbf{0.84} & \textbf{0.34} & \textbf{0.22} & \textbf{0.71} & \textbf{0.57} & -- & -- \\
\midrule
\multirow{4}{*}{Remove}
 & M2UGen            & 3.00 & 0.27 & 1.28 & 0.09 & 0.36 & $-46.01$ & $-52.43$
                     & 4.49 & 0.28 & 1.37 & 0.09 & 0.27 & $-43.21$ & $-49.87$ \\
 & Instruct-MG       & 1.42 & \textbf{0.37} & 0.50 & 0.43 & 0.55 & $-2.10$ & $-8.53$
                     & 3.55 & 0.20 & 0.71 & 0.37 & 0.49 & $-4.41$ & $-11.03$ \\
 & LeVo              & 2.20 & 0.35 & 0.98 & 0.10 & 0.31 & $-40.32$ & $-46.74$
                     & 3.50 & 0.31 & 1.01 & 0.10 & 0.26 & $-41.85$ & $-48.47$ \\
 & \textbf{AURA}     & \textbf{0.34} & 0.36 & \textbf{0.12} & \textbf{0.80} & \textbf{0.74} & $\mathbf{+11.32}$ & $\mathbf{+4.89}$
                     & \textbf{0.72} & \textbf{0.33} & \textbf{0.22} & \textbf{0.70} & \textbf{0.78} & $\mathbf{+9.16}$ & $\mathbf{+2.54}$ \\
\midrule
\multirow{4}{*}{Extract}
 & M2UGen            & 6.66 & 0.32 & 1.36 & 0.16 & 0.67 & $-45.47$ & $-39.32$
                     & 4.26 & 0.18 & 1.46 & 0.14 & 0.65 & $-43.61$ & $-39.95$ \\
 & Instruct-MG       & 5.32 & \textbf{0.44} & \textbf{0.74} & 0.27 & 0.79 & $-15.18$ & $-9.05$
                     & 5.20 & 0.18 & \textbf{0.72} & 0.28 & 0.79 & $-14.76$ & $-10.39$ \\
 & LeVo              & 5.95 & 0.29 & 1.76 & 0.12 & 0.44 & $-39.80$ & $-33.67$
                     & 4.44 & 0.17 & 1.41 & 0.09 & 0.40 & $-40.49$ & $-36.11$ \\
 & \textbf{AURA}     & \textbf{4.48} & 0.41 & 0.86 & \textbf{0.45} & \textbf{0.80} & $\mathbf{-7.62}$ & $\mathbf{-1.49}$
                     & \textbf{4.24} & \textbf{0.26} & \textbf{0.72} & \textbf{0.48} & \textbf{0.89} & $\mathbf{+2.67}$ & $\mathbf{+7.04}$ \\
\bottomrule
\end{tabular}
\vspace{-2mm}
\end{table*}
\noindent\textbf{Training dataset.}
We train on \textbf{66{,}539} conversational dialogues generated from the Slakh2100 training subset. Each dialogue contains a user turn (source audio + LLM-paraphrased instruction) with an assistant
turn containing a typed edit-token block over 10\,s EnCodec windows. Every
sample is annotated with edit kind, a 10-class instrument label (94.5\%
coverage). For evaluation, we use 1000
in-domain samples from the  Slakh test split (add/remove/extract
at 32\,kHz) and 1000 out-of-domain samples from MoisesDB with
LLM-written instructions.

\noindent\textbf{Implementation details.} The MLLM emits \(n_c=9\) concept tokens, which condition the frozen MusicGen backbone with hidden size \(d_m=1536\). AdamW is used with a learning rate of \(10^{-4}\), an effective batch size of 16, and rank-64 LoRA (\(\alpha=128\)) on the cross-attention key and value projections. Inference uses classifier-free guidance with a scale of 2.0.

\noindent\textbf{Evaluation.}
We evaluate methods along four criteria: audio quality (FAD$\downarrow$~\cite{kilgour2018fr}, KL$\downarrow$), instruction adherence (CLAP$\uparrow$~\cite{elizalde2023clap}), edit correctness (P-Demucs$\uparrow$~\cite{defossez2019demucs}), and content preservation (SSIM$\uparrow$, SI-SDR$\uparrow$, and SI-SDRi$\uparrow$)~\cite{le2019sdr}. SI-SDR and SI-SDRi are reported only for removal and extraction tasks, where deterministic targets are available. In our evaluation protocol, we conduct single-turn editing and multi-turn editing to assess the effectiveness of AURA.

\vspace{-2mm}
\subsection{Quantitative results}

\noindent\textbf{Single-turn editing.} Table~\ref{tab:baselines} shows that AURA consistently improves edit correctness and content preservation. On Slakh2100, it more than doubles the best baseline SSIM for \textit{add} (0.78 vs.\ 0.37) and uniquely achieves positive SI-SDR for \textit{remove} (\(+11.32\,\mathrm{dB}\)). These gains generalize to MoisesDB, where AURA reduces FAD by \(4\)--\(5\times\) for \textit{add} and \textit{remove} and achieves positive SI-SDR for \textit{extract}. Although its in-domain CLAP and extraction quality remain less consistent, AURA substantially improves preservation across tasks and domains.

\begin{table}[t]
\footnotesize\centering
\setlength{\tabcolsep}{4pt}
\caption{Multi-turn editing benchmark MoisesDB (out-of-domain). Turns apply
$-$drums, $+$bass, $+$guitar in sequence; each is scored against the exact
intermediate target. The control (Ctrl) applies only the final edit to the
original mix, without the intermediate turns.}
\label{tab:multiturn}
\begin{tabular}{ll|cccc}
\toprule
\textbf{Metric} & \textbf{Model} & \textbf{T1} & \textbf{T2} & \textbf{T3} & \textbf{Ctrl} \\
\midrule
\multirow{4}{*}{SSIM$\uparrow$}
 & M2UGen      & 0.10 & 0.08 & 0.10 & 0.11 \\
 & Instruct-MG & 0.42 & 0.32 & 0.22 & 0.32 \\
 & LeVo        & 0.08 & 0.07 & 0.08 & 0.08 \\
 & \textbf{AURA} & \textbf{0.84} & \textbf{0.79} & \textbf{0.54} & \textbf{0.49} \\
\midrule
\multirow{4}{*}{SI-SDR$\uparrow$}
 & M2UGen      & $-44.8$ & $-45.9$ & $-46.1$ & $-47.1$ \\
 & Instruct-MG & $-0.9$  & $-8.7$  & $-19.3$ & $-13.7$ \\
 & LeVo        & $-38.7$ & $-49.8$ & $-52.7$ & $-51.0$ \\
 & \textbf{AURA} & $\mathbf{+21.3}$ & $\mathbf{+12.3}$ & $\mathbf{+5.1}$ & $-5.2$ \\
\midrule
\multirow{4}{*}{SI-SDRi$\uparrow$}
 & M2UGen      & $-59.6$ & $-48.3$ & $-43.3$ & $-44.4$ \\
 & Instruct-MG & $-15.7$ & $-11.1$ & $-16.5$ & $-10.9$ \\
 & LeVo        & $-47.3$ & $-47.8$ & $-46.6$ & $-45.0$ \\
 & \textbf{AURA} & $\mathbf{+6.5}$ & $\mathbf{+9.9}$ & $\mathbf{+7.9}$ & $-2.4$ \\
\bottomrule
\end{tabular}
\vspace{-4mm}
\end{table}

\noindent\textbf{Multi-turn editing.}
Table~\ref{tab:multiturn} evaluates all methods under the same cascaded
protocol. At each turn, M$^2$UGen, LeVo, and Instruct-MG receive their own
previous output as the new reference audio together with the current
instruction, while AURA additionally receives the complete dialogue
history. AURA maintains positive SI-SDR from \(+21.3\) to
\(+5.1\,\mathrm{dB}\) and positive SI-SDRi across all turns. In contrast,
the strongest baseline, Instruct-MG, degrades from \(-0.9\) to
\(-19.3\,\mathrm{dB}\), demonstrating greater error accumulation under
sequential editing. Applying only the final instruction to the original
mixture reduces AURA's SI-SDR to \(-5.2\,\mathrm{dB}\), indicating that
the dialogue history is necessary to recover the intended editing state.
Although AURA's SSIM decreases from 0.84 to 0.54, it remains substantially
more robust than the single-turn baselines.


\vspace{-2mm}
\subsection{Ablation study}
Table~\ref{tab:ablation} compares concept-token fusion with the decoder's original full-text conditioning. Concept tokens consistently improve audio quality, suggesting that their compact semantic representations align more effectively with the music decoder and are less susceptible to irrelevant variation in the full-sentence hidden states. These results validate the importance of concept-token-based fusion.
\begin{table}[t]
\footnotesize\centering
\setlength{\tabcolsep}{4pt}
\caption{Comparison of the effects of $\langle\text{EDIT\_CONCEPT}\rangle$ tokens with fully LLM-generated token hidden states on the MoiseDB dataset. Best results are in \textbf{bold}.}
\label{tab:ablation}
\begin{tabular}{l|cc|cc|cc}
\toprule
& \multicolumn{2}{c|}{FAD$\downarrow$} & \multicolumn{2}{c|}{CLAP$\uparrow$} & \multicolumn{2}{c}{SSIM$\uparrow$} \\
\cmidrule(lr){2-3}\cmidrule(lr){4-5}\cmidrule(lr){6-7}
\textbf{Task} & Text & Concept & Text & Concept & Text & Concept \\
\midrule
Add     & 13.32 & \textbf{0.84} & 0.10 & \textbf{0.34} & 0.08 & \textbf{0.71} \\
Remove  & 9.49  & \textbf{0.72} & 0.15 & \textbf{0.33} & 0.068 & \textbf{0.70} \\
Extract & 9.97  & \textbf{4.24} & 0.14 & \textbf{0.26} & 0.099 & \textbf{0.48} \\
\bottomrule
\end{tabular}
\vspace{-4mm}
\end{table}

%% file: sections/4conclusion.tex
\vspace{-3mm}
\section{Conclusion}
\vspace{-2mm}
We introduced AURA, a unified multimodal framework that brings
conversational control to instruction-guided music editing. By routing all
multimodal and dialogue context through a compact set of concept tokens,
and by injecting the reference audio frame-aligned rather than as a style
prompt, AURA separates \emph{understanding what to change} from
\emph{producing the result} while keeping the pretrained music decoder
frozen. Experiments on in-domain and out-of-domain benchmarks show large
gains in content preservation and edit correctness over prior editors. The
main limitation is stem extraction, where a generative decoder cannot
match dedicated separation models; coupling AURA with an explicit
separation front-end is a natural next step. 
